\documentclass[aps,prb,twocolumn,showpacs,superscriptaddress,amsmath,amssymb,floatfix,10pt]{revtex4-1}  
\usepackage{graphicx}  
\usepackage{dcolumn}   
\usepackage{bm}        
\usepackage{amssymb}   
\usepackage{framed}
\usepackage{amsmath}
\usepackage{multirow}
\usepackage{hhline}
\usepackage{color}
\definecolor{bluegreen}{rgb}{0,0.2,0.8}
\usepackage{subfigure,amsmath,verbatim,moreverb}
\usepackage{tabularx}
\usepackage{adjustbox}
\usepackage{lipsum}
\usepackage{longtable}
\usepackage{booktabs}

\usepackage{etoolbox}
\AtBeginEnvironment{align}{\setcounter{subeqn}{0}}
\newcounter{subeqn} %

\begin{document}

\title{Assessing the performance of the recent non-empirical semilocal density functionals on
describing the lattice constants, bulk moduli and cohesive energies of alkali, alkaline-earth, and
transition metals}
\author{Subrata Jana}
\email{subrata.jana@niser.ac.in}
\affiliation{School of Physical Sciences, National Institute of Science Education and Research, HBNI, 
Bhubaneswar 752050, India}
\author{Kedar Sharma}
\affiliation{School of Physics, Indian Institute of Science Education and Research, Maruthamala, Vithura, Thiruvananthapuram 695551, India}
\author{Prasanjit Samal}
\affiliation{School of Physical Sciences, National Institute of Science Education and Research, HBNI, 
Bhubaneswar 752050, India}

\date{\today}

\begin{abstract}

The bulk properties (lattice constants, bulk moduli, and cohesive energies) of alkali,
alkaline-earth, and transition metals are studied within the framework of the recently 
developed meta-GGA (meta-Generalized Gradient Approximation) semilocal exchange-correlation
functionals. To establish the applicability, broadness and accuracy of meta-GGA functionals
we also put the results of PBE (Perdew-Burke-Ernzerhof) and PBEsol (PBE reparametrized for
solids) functionals. The interesting feature of the present paper is that it measures the
accuracy of the recently developed TM (Tao-Mo) and TMTPSS (TM exchange with 
Tao-Perdew-Staroverov-Scuseria (TPSS) correlation) and  SCAN (Strongly Constrained and
Appropriately Normed) functionals on describing aforementioned properties. The present
systematic investigation shows that the TM is accurate in describing the lattice constants
while for cohesive energies and bulk moduli the accuracy is biased towards the PBE and TPSS
functionals.

\end{abstract}

\maketitle

\section{Introduction}

Since its advent, the Kohn-Sham (KS)~\cite{KS65} density functional theory (DFT) is a {\it
de facto} standard theoretical framework for studying the electronic structures of solids and
materials. The accuracy of the KS density functional depends on the accuracy of the
approximate exchange-correlation functionals. Due to the reasonable computational cost with
comparatively well balanced accuracy the semilocal nature of the exchange-correlation
functionals i.e., the local density approximations (LDA)~\cite{lda}, generalized-gradient
approximation (GGA)~\cite{PW86,B88,LYP88,PW91,B3PW91,PBE96,AE05,ZWu06,SOGGA,PBEsol,con1,
con2,con3,con4} and meta-generalized gradient approximation (meta-GGA)~\cite{BR89,
VSXC98,MO6L,TPSS03,revTPSS,con5,MS0,MS1MS2,SCAN15,Tao-Mo16} are widely used for the bulk
properties of solids~\cite{YMo16,YMoCPL,mapkw,smrkp,mgga-vasp,perdewpnas,PJanthon13,
PJanthon14,PBlaha09,FTran16,GCsonks09,GKresse13,SJana18}. The mainstream of semilocal
density functionals are developed based on the constraint satisfaction~\cite{PBE96,
AE05,ZWu06,SOGGA,PBEsol,con1,con2,con3,con4,TPSS03,revTPSS,con5,MS0,MS1MS2,SCAN15,Tao-Mo16}
or modelling the exchange hole~\cite{Tao-Mo16} or association both the properties
~\cite{Tao-Mo16}. Usually, benchmarking the density functional approximations against the
experimental features are common practice to measure the accuracy and applicability of the
approximation, in particular when a new functional is introduced. Indeed, the systematic
evaluation of the properties of a density functional approximation guide users to properly
choose a functional for describing the material properties. Also, the behavior of the
functionals for the wide range of systems makes it easier to improve the drawback of the
functional.

The present paper seeks to assess the performance of the recently proposed meta-GGA
functionals at the accuracy of the bulk properties of the transition metals. More
specifically, we consider the lattice constants (or equilibrium shortest distances), bulk
moduli, and cohesive energies of transition metals. Regarding the performance of different
level of density functional approximations for the bulk properties of transition metals, it
has been studied earlier within the framework of  GGA, meta-GGA and hybrid functionals
theory. In ref.~\cite{PJanthon13} Janthon et. al. studied the transition metals within the
framework of GGA based functionals. In ref.~\cite{PJanthon14} Janthon et. al. further
explore the behavior of transition metals by including meta-GGA level functionals. Besides,
Hass et. al.~\cite{PBlaha09}, Csonka et. al.~\cite{GCsonks09}, Tran et. al.~\cite{FTran16},
Schimka et. al.~\cite{GKresse13}, Hao et. al.~\cite{PHao12}, and Zhang et. al.~\cite{GZhang08} partially cover the lattice constants, cohesive
energies and bulk moduli of alkali, alkaline-earth, and transition metals. In this paper,
we put together the lattice constants, cohesive energies and bulk moduli of all the alkali,
alkaline-earth, and transition metals within the framework of recently developed meta-GGA
functionals. Our comparison meta-GGA functionals contain Tao-Perdew-Staroverov-Scuseria
(TPSS)~\cite{TPSS03}, revised TPSS  (revTPSS)~\cite{revTPSS}, Minnesota 2006 local
functional (M06L)~\cite{MO6L}, meta-GGA made simple (MS0, MS1, and
MS2)~\cite{MS0,MS1MS2,MSh} functionals, Strongly Constrained and Appropriately Normed
(SCAN)~\cite{SCAN15}, and Tao-Mo~\cite{Tao-Mo16} meta-GGA functional (TMTPSS and TM)
functionals.

Arguably, the recent advances in the meta-GGA functionals show that the accuracy of the
functionals can be further improved by imposing more exact constraints on the functional
construction. The motivation of the present paper flows from the appealing features and
accuracy of the recently developed meta-GGA functionals. The recent development of the
meta-GGA functionals shows that the SCAN functionals developed by Sun et. al.~\cite{SCAN15}
and TM functional developed by Tao et. al.~\cite{Tao-Mo16} quite accurate in describing
several bulk properties of solids. Though the SCAN functional has been studied for bulk
properties of few metals but remains untested extensively for the alkali, alkaline-earth,
and transition metals. Also, the recently developed TM functional remains untested for
those properties. In the present paper, we present the benchmark calculations of the bulk
properties of the $3d$, $4d$ and $5d$ transition metals, alkali, and alkaline earth metals.
In particular, due to different bonding nature of the alkali, alkaline-earth and transition
metals, they are considered as the difficult cases within semilocal exchange-correlation
functionals. Though they are considered mainly as metals and metallic bonding dominates,
but the weak van-der-Waals interactions in closed semi-core states also play important
role~\cite{GKresse13, JTao10}. These make the semilocal functionals difficult to describe
accurately all the bonding nature and often in the benchmarking calculations the $3d$, $4d$
and $5d$ transition metals, alkali and alkaline earth metals are excluded.

It was shown in ref.~\cite{GKresse13} that the  PBE functional does not perform in a
satisfactory way in describing the lattice constants of all these metals. It was shown that
the PBE lattice constants for $3d$ metals are slightly too small, whereas, the lattice
constants reported for $4d$ and $5d$ metals using PBE are too large. It was also shown in
ref.~\cite{GKresse13} that the situation improves through the inclusion of kinetic-energy
density term in the functional construction. Due to the one electron free correlation of
the revTPSS functional, it performs reasonably for the transition metal lattice constants.
All these previous studies motivate us to assess the accuracy of the recently developed
meta-GGAs in predicting the aforementioned bulk properties. It is noteworthy to mention
that the SCAN functional includes the intermediate van-der-Waals (vdW) interaction,
therefore it will be an interesting study to assess its performance for alkali and
alkaline-earth materials, where the bonding is influenced by the vdW interaction in the
semi-core states. Regarding the TM functionals, it was shown that both the TM and TPSS
correlation perform differently with the TM exchange. But the accuracy of the TMTPSS and TM
has not been measured against the $3d$, $4d$ and $5d$ transition metals, alkali, and
alkaline earth metals. In this paper, we put all the modern meta-GGA density functionals
(TPSS, revTPSS, M06L, MS0, MS1, MS2, SCAN, TMTPSS, and TM) together with GGA
(PBE~\cite{PBE96} and PBEsol~\cite{PBEsol}) based semilocal functionals to assess the
performance of alkali, alkaline-earth, and transition metals.

This paper is organized as follows: In the following, we will describe our computational
set up along with the test set used for our calculations. Following that, we will study the
lattice constants, bulk moduli and surface energies of the transition metals. We will
conclude by discussing and comparing our results.

\section{Computational Setup}

All computational studies are performed using the plane wave code based on the
projector-augmented method Vienna {\it Ab Initio} Simulation Package (VASP)~\cite{paw1,
paw2,vasp1,vasp2,vasp3,uspp}. The Bulk calculations are performed with $16\times 16\times
16$ Gamma-centered $\bf{k}-$ points. Regarding the atomic calculations of cohesive energies
a simulation box of $20\times 20 \times 20$~\AA$^3$ ~has been used with $1\times 1\times 1$
Gamma-centered $\bf{k}-$ points. The spin polarization calculations are performed for
atoms. An energy convergence criterion of $10^{-6}$ has been set for bulk calculations,
whereas, the atomic simulations are performed with energy convergence criterion of
$10^{-5}$. It is noteworthy to mention that all calculations of the meta-GGA functionals
are performed by starting from the PBE wavefunctions and change densities. The energy
cutoff $500$ eV to $700$ eV is used for bulk calculations, whereas $700$ eV to $1000$ eV
energy cutoff is used for the atomic calculations.

The results present in TABLEs are arranged by separation out the $3d$ transition metals,
$4d$ transition metals, $5d$ transition metals, alkali metals (K, Rb, and Cs) and
alkaline-earth metals (Ca, Sr and Ba). Under the ambient condition, all the alkali metals,
alkaline-earth metals, and transition metals show $fcc$, $bcc$, or $hcp$ structures. Only
exceptions are Mn, La, and Hg. These materials show complicated hexagonal (La),
rhombohedral (Hg) and cubic unit cell with 58 atoms (Mn). Due to the very different
structures of La, Hg, and Mn, they are also discussed separately in all TABLEs.

To compare the overall accuracy of all the functionals we present mean-error (ME), mean
absolute error (MAE) of the $3d$, $4d$, $5d$ along with alkali and alkaline metals. The
total ME and total MAE is also given in the last column of each TABLE. 

 \section{Results and Discussions}
 
 \subsection{Equilibrium Inter-atomic Shortest Distances}
 
All the lattice constants of alkali, alkaline-earth, and transition metals are determined
at their ambient conditions and non-magnetic phases. The experimental values presented in
TABLE I are subtracted for the zero-point vibrational effects (ZPVE). TABLE I presents the
ZPVE corrected equilibrium inter-atomic distances along with the benchmark results of all
the functionals. The equilibrium inter-atomic distances depend on the equilibrium lattice
constants according to the different lattice structures. For the details of the relation between 
the inter-atomic distances and equilibrium lattice constants, the readers are suggested to go
through the ref.~\cite{PJanthon13}.  In FIG. I, we also show the percentage deviation of
our calculated values. Though the calculations for the PBE, PBEsol, M06L, TPSS, and revTPSS
are reported in past for several solids, but in our present study, we
recalculate all the solids along with recently developed meta-GGAs. In this section, we
will discuss the functional performances according to the data present in TABLE I. 

 \begin{longtable*}{lcccccccccccccccc}
\label{latt-cons}\\
 \caption{Equilibrium shortest distances $\delta$ (in Picometre (pm)) of different solid structures using
 PBE, PBEsol, TPSS, revTPSS, M06L, MS0, MS1, MS2, SCAN, TMTPSS and TM functionals. 
 The experimental reference values are collected from references~\cite{PJanthon14,GKresse13,
 GCsonks09}, where the correction due to the zero-point vibrational effects (ZPVE) are taken into account. 
 For elements Mn, La and Hg the ZPVE corrected values are not available and we reported only the 
 experimental values taken from reference~\cite{PJanthon13}.}\\
\hline\hline
Metals&PBE&PBEsol&TPSS&revTPSS&M06L&MS0&MS1&MS2&SCAN&TMTPSS&TM&Expt.\\
\hline\hline 
Sc	&	330.1	&	326.3	&	328.4	&	327.7	&	328.0	&	331.2	&	331.5	&	328.9	&	329.6	&	328.7	&	327.8	&	324.4	\\
Ti	&	292.3	&	288.8	&	290.4	&	289.2	&	291.0	&	291.9	&	292.1	&	290.4	&	290.8	&	291.2	&	290.1	&	288.9	\\
V	&	258.0	&	254.4	&	256.1	&	255.2	&	257.3	&	255.5	&	255.8	&	255.2	&	255.6	&	256.8	&	256.1	&	260.6	\\
Cr	&	245.6	&	242.4	&	244.0	&	243.0	&	244.3	&	242.7	&	243.0	&	242.7	&	243.2	&	244.3	&	243.5	&	248.5	\\
Fe	&	238.6	&	234.9	&	236.6	&	235.5	&	236.3	&	235.0	&	235.3	&	235.1	&	235.1	&	236.8	&	236.0	&	245.0	\\
Co	&	245.1	&	240.9	&	242.9	&	241.6	&	243.0	&	240.9	&	241.3	&	241.1	&	241.2	&	243.0	&	242.0	&	248.8	\\
Ni	&	248.2	&	243.6	&	245.2	&	243.2	&	241.8	&	243.8	&	244.2	&	243.7	&	243.7	&	244.9	&	243.6	&	248.4	\\
Cu	&	257.0	&	251.5	&	253.0	&	250.9	&	248.3	&	250.5	&	251.3	&	250.2	&	249.4	&	252.1	&	251.1	&	254.4	\\
Zn	&	263.5	&	261.1	&	263.2	&	261.4	&	263.4	&	260.5	&	260.9	&	259.7	&	258.8	&	261.0	&	260.5	&	264.5	\\
	&		&		&		&		&		&		&		&		&		&		&		&		\\
ME	&	-0.6	&	-4.4	&	-2.6	&	-4.0	&	-3.3	&	-3.5	&	-3.1	&	-4.1	&	-4.0	&	-2.7	&	-3.6	&		\\
MAE	&	3.2	&	4.8	&	3.9	&	4.8	&	4.6	&	5.7	&	5.4	&	5.4	&	5.6	&	4.2	&	4.7	&		\\
	&		&		&		&		&		&		&		&		&		&		&		&		\\
Y	&	363.3	&	358.5	&	363.8	&	361.6	&	366.6	&	366.7	&	366.6	&	363.9	&	365.7	&	365.5	&	362.6	&	354.8	\\
Zr	&	323.6	&	318.0	&	321.4	&	319.2	&	324.0	&	321.4	&	321.5	&	320.5	&	321.1	&	323.1	&	320.2	&	317.4	\\
Nb	&	287.8	&	284.1	&	286.7	&	285.2	&	288.2	&	285.5	&	285.7	&	285.6	&	286.4	&	287.2	&	286.1	&	285.4	\\
Mo	&	272.9	&	269.9	&	271.9	&	270.3	&	271.9	&	270.2	&	270.4	&	270.4	&	271.1	&	271.7	&	270.6	&	272.1	\\
Tc	&	274.4	&	271.2	&	273.1	&	271.3	&	272.2	&	271.0	&	271.2	&	271.3	&	272.0	&	272.8	&	271.5	&	270.5	\\
Ru	&	271.6	&	268.1	&	270.5	&	268.1	&	269.0	&	266.9	&	267.2	&	267.7	&	267.4	&	270.0	&	268.3	&	264.2	\\
Rh	&	270.4	&	266.2	&	268.9	&	263.6	&	264.2	&	263.0	&	263.3	&	263.5	&	263.8	&	264.3	&	263.9	&	253.2	\\
Pd	&	278.6	&	273.2	&	276.1	&	272.9	&	277.2	&	272.8	&	273.2	&	273.0	&	273.8	&	275.9	&	273.6	&	274.5	\\
Ag	&	293.4	&	285.7	&	289.1	&	285.0	&	291.4	&	285.7	&	286.2	&	285.3	&	286.1	&	288.4	&	285.6	&	287.7	\\
Cd	&	302.0	&	306.2	&	301.3	&	298.4	&	311.8	&	313.9	&	298.2	&	297.8	&	296.3	&	299.6	&	298.0	&	295.9	\\
	&		&		&		&		&		&		&		&		&		&		&		&		\\
ME	&	6.2	&	2.5	&	4.7	&	2.0	&	6.1	&	4.1	&	2.8	&	2.3	&	2.8	&	4.3	&	2.5	&		\\
MAE	&	6.2	&	3.9	&	4.8	&	3.3	&	6.1	&	5.3	&	3.7	&	3.5	&	3.5	&	4.4	&	3.4	&		\\
	&		&		&		&		&		&		&		&		&		&		&		&		\\
Hf	&	319.5	&	315.2	&	317.1	&	315.3	&	320.4	&	316.1	&	316.5	&	315.4	&	315.1	&	316.6	&	315.9	&	312.6	\\
Ta	&	286.6	&	283.3	&	284.8	&	283.1	&	287.2	&	282.9	&	283.2	&	282.9	&	282.9	&	284.4	&	283.7	&	285.6	\\
W	&	274.7	&	272.0	&	273.3	&	271.7	&	273.5	&	271.4	&	271.7	&	271.6	&	271.8	&	272.8	&	272.1	&	273.8	\\
Re	&	277.2	&	274.4	&	273.4	&	272.5	&	273.0	&	271.4	&	271.8	&	272.2	&	271.1	&	273.2	&	272.8	&	256.2	\\
Os	&	275.4	&	272.4	&	274.1	&	271.8	&	272.5	&	270.9	&	271.2	&	271.6	&	270.6	&	273.2	&	271.9	&	267.1	\\
Ir	&	273.8	&	270.5	&	272.6	&	270.1	&	271.4	&	268.5	&	268.9	&	269.6	&	267.6	&	271.7	&	270.2	&	271.0	\\
Pt	&	280.5	&	276.1	&	278.9	&	275.6	&	278.8	&	274.3	&	274.6	&	275.1	&	274.8	&	278.0	&	275.8	&	276.6	\\
Au	&	293.9	&	287.7	&	291.1	&	286.9	&	291.6	&	285.8	&	286.3	&	286.4	&	286.8	&	290.3	&	287.2	&	287.0	\\
	&		&		&		&		&		&		&		&		&		&		&		&		\\
ME	&	6.5	&	2.7	&	4.4	&	2.1	&	4.8	&	1.4	&	1.8	&	1.9	&	1.4	&	3.8	&	2.5	&		\\
MAE	&	6.5	&	4.0	&	4.8	&	3.8	&	4.9	&	4.2	&	4.1	&	4.0	&	3.9	&	4.3	&	3.8	&		\\
	&		&		&		&		&		&		&		&		&		&		&		&		\\
K	&	457.5	&	452.5	&	463.5	&	459.0	&	427.8	&	463.5	&	464.0	&	459.2	&	458.9	&	446.1	&	446.0	&	451.4	\\
Rb	&	490.1	&	483.7	&	497.1	&	492.6	&	451.6	&	499.6	&	502.4	&	493.7	&	491.7	&	483.7	&	483.7	&	483.0	\\
Cs	&	534.8	&	520.3	&	542	&	536.5	&	482.2	&	546.9	&	549.3	&	538.6	&	539.9	&	523.5	&	523.5	&	523.0	\\
	&		&		&		&		&		&		&		&		&		&		&		&		\\
ME	&	8.3	&	-0.3	&	15.1	&	10.2	&	-31.9	&	17.5	&	19.4	&	11.4	&	11.0	&	-1.4	&	-1.4	&		\\
MAE	&	8.3	&	1.5	&	15.1	&	10.2	&	31.9	&	17.5	&	19.4	&	11.4	&	11.0	&	2.2	&	2.2	&		\\
	&		&		&		&		&		&		&		&		&		&		&		&		\\
Ca	&	394.3	&	385.8	&	391.1	&	390.0	&	378.7	&	394.3	&	395.5	&	392.7	&	393.5	&	388.6	&	388.1	&	392.9	\\
Sr	&	426.4	&	418.7	&	426.6	&	425.1	&	414.9	&	431.2	&	433.2	&	427.3	&	430.3	&	424.2	&	423.3	&	427.1	\\
Ba	&	423.7	&	423.6	&	433.5	&	431.5	&	430.8	&	440.2	&	441.6	&	435.9	&	435.9	&	433.5	&	431.2	&	433.2	\\
	&		&		&		&		&		&		&		&		&		&		&		&		\\
ME	&	-2.9	&	-8.4	&	-0.7	&	-2.2	&	-9.6	&	4.2	&	5.7	&	0.9	&	2.2	&	-2.3	&	-3.5	&		\\
MAE	&	3.9	&	8.4	&	0.9	&	2.2	&	9.6	&	4.2	&	5.7	&	1.0	&	2.2	&	2.5	&	3.5	&		\\
	&		&		&		&		&		&		&		&		&		&		&		&		\\
Mn	&	230.7	&	227.1	&	228.4	&	227.7	&	228.6	&	227.4	&	227.7	&	227.4	&	227.6	&	228.6	&	230.2	&	224.0	\\
La	&	376.9	&	365.4	&	373.3	&	369.2	&	385.9	&	373.9	&	374.7	&	372.1	&	379.5	&	376.6	&	373.4	&	373.9	\\
Hg	&	323.8	&	300.6	&	308.5	&	299.5	&	308.9	&	297.6	&	298.6	&	297.6	&	300.2	&	305.2	&	299.5	&	301.0	\\
	&		&		&		&		&		&		&		&		&		&		&		&		\\
ME	&	10.8	&	-1.9	&	3.8	&	-0.8	&	8.2	&	0.0	&	0.7	&	-0.6	&	2.8	&	3.8	&	1.4	&		\\
MAE	&	10.8	&	4.0  	&	4.2	&	3.3	   &	8.2	&	2.3	&	2.3	&	2.9  	&	3.3	&	3.8	&	2.7	&		\\
	&		&		&		&		&		&		&		&		&		&		&		&		\\

	\hline\hline
TME	   &	4.4	&	-0.7	&	3.1	&	0.6	&	-0.9	&	2.4	&	2.5	&	1.0	&	1.3	&	1.4	&	0.0	&		\\
TMAE	&	5.9	&	4.3	   &	5.0	&	4.3	&	8.1	    &	5.8	&	5.6	&	4.5	&	4.5	&	3.9	&	3.6	&		\\
\hline
\hline

\end{longtable*}

 \begin{figure*}
\includegraphics[width=7.0in,height=5.0in,angle=0.0]{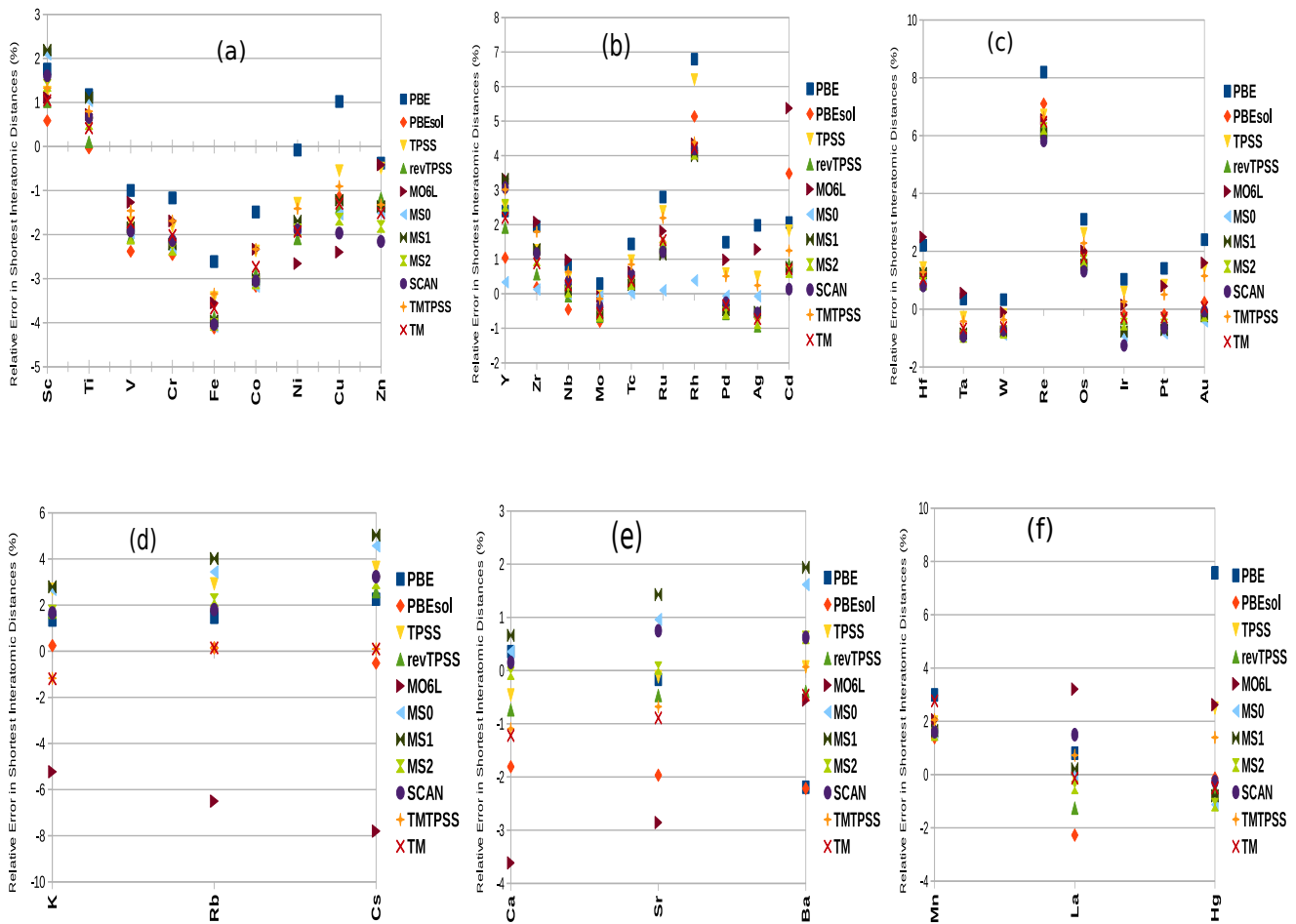}
\caption{Histograms of relative error in the interatomic shortest distances. The percentage deviation 
is plotted along Y-axis. The numbering of the figures are as the order of the solids presented in TABLE I.}
\label{lattice-plot}
\end{figure*}

 \textbf{$3d$ transition metals :}~ The inter-atomic distances of $3d$ elements are
 presented at the top of the TABLE I. The $3d$ elements contain with Sc, Ti, V, Cr, Fe,
 Co, Ni, Cu, and Zn. Let us start our discussion with the popularly used GGA functional
 PBE. The PBE functional perform fairly good throughout the series. However, It
 overestimates the inter-atomic distances of Sc, Ti, V, and Cu, and underestimations the
 inter-atomic distances for Cr, Fe, and Co. But it gives a fairly good inter-atomic
 distance for Ni. Regarding the performance of the PBEsol, it underestimates the
 inter-atomic distances of all the elements except Sc and Ti. For Sc, the PBEsol
 overestimates the inter-atomic distance, whereas, very good inter-atomic distance is
 obtained for Ti. The underestimation percentage of all the elements are fairly large for
 PBEsol compared to PBE. Especially, for Fe, a fairly large underestimation is observed. Now,
 we consider the meta-GGA functionals. Regarding the performance for the TPSS and revTPSS
 functionals, the revTPSS functional lower the inter-atomic distances compared to TPSS for
 all the $3d$ elements by almost $1$ pm to $3$ pm. TPSS overestimates the inter-atomic
 distances for Sc, Ti, but it follows the different trend as the $d$ bands become filled.
 For V, Cr, Fe, Co, Ni, Cu and Zn TPSS underestimates the inter-atomic distances. The
 underestimation tendency becomes less intense as the $d$ band almost filled. Now,
 concerning the performance of M06L, it follows the same trend as TPSS does except for Ni
 and Cu. For Ni and Cu, M06L underestimates the inter-atomic distances more than TPSS and
 revTPSS do. Now, for the meta-GGA made simple functionals (MS0, MS1, and MS2), all perform
 equivalently and overestimate the inter-atomic distances for Sc and Ti, but underestimate
 the inter-atomic distances form V to Zn. The underestimation and overestimation percentage
 of all the ``MS'' functionals are fairly large compared to the traditional TPSS and
 revTPSS functionals. Now, we consider the performance of the two recently proposed
 functionals SCAN and TM (TMTPSS and TM). The performance of SCAN quite disappointing as it
 follows the same trend of ``MS'' functionals. The MAE of SCAN functional indicates that it
 performs better than MS0 but less accurate than MS1 and MS2. Now, concerning TMTPSS and TM
 functionals both are less accurate than TPSS in predicting the inter-atomic distances of
 $3d$ metals. In fact, both show the same tradition as it is observed using other
 meta-GGAs. Overall consideration shows that the PBE functionals perform fairly well for
 all the $3d$ elements and reports the best MAE compared to the more advanced meta-GGA
 functionals. Overall tendency of meta-GGAs shows that TPSS is the best among all the
 meta-GGA functionals. We do not observe improvement in the inter-atomic distances using
 the more advance functionals like SCAN and TM based functionals.

 \textbf{$4d$ transition metals :}~Unlike the $3d$ elements, in this case, the PBE
 functional overestimates the the inter-atomic distances of all the elements except Mo. A
 sizable overestimation in the inter-atomic distances is observed for Y, Zr, Tc, Ru, Rh, Ag
 and Cd. For Nb, the overestimation is observed within the limit of $\approx 2$ pm. The
 PBEsol functional reduce the sizable overestimation percentage of PBE. Concerning the
 meta-GGAs functionals, very good MAE is observed using revTPSS, MS1, MS2, SCAN and TM
 functionals. A sizable overestimation in the interatomic distances is observed using the
 TPSS, M06L and MS0 functionals. Among all the meta-GGAs, the error obtained using the M06L
 functional is fairly large. It actually follows PBE like tendency in this case. Regarding
 the recent two meta-GGAs, SCAN and TM, both perform equivalently for all the elements
 except Y. In that case, the SCAN overestimates more than TM functionals. For $4d$
 transition metals, the TM functional performs better than TMTPSS functionals. For $4d$
 transition metal elements, all semilocal functionals show overestimation tendency in
 predicting the inter-atomic distances except for few cases.

 \textbf{$5d$ transition metals :}~We observe the same trend as it is observed for the $4d$
 elements. A fairly sizable overestimation is observed for PBE functional. PBEsol reduces
 the MAE for the PBE functional. In case of meta-GGA functionals, the revTPSS, SCAN, and TM
 functionals perform equivalently. Using the ``MS'' meta-GGAs we obtain almost equivalent
 MAE. Interestingly, the ``MS'' meta-GGA functionals overall overestimate the interatomic
 distances for few cases, whereas, underestimation in the inter-atomic distances are
 observed for others. A similar tendency is observed for the SACN functional. Both the TM
 and TMTPSS functionals quite reasonably predict the inter-atomic distances for the Ta, W,
 Os, Ir, Pt, and Au. For other elements overestimation is observed. Overall we obtain least
 MAE using the revTPSS and TM functionals.      
 
 \textbf{Alkali metals :}~The alkali metals contain elements K, Rb and Cs. We separately
 discuss these metals because different kinds of interactions affect the bonding of these
 metals. Though they are considered as prototypical metals but a contribution from the
 semi-core $p$ and $s$ states originate van-der-Waals bonding~\cite{GKresse13, JTao10}
 which affect the lattice constants or equilibrium shortest distances. A sizable error in
 equilibrium shortest distances is observed from the PBE functionals. The PBE functionals
 overestimate the equilibrium shortest distances for all the alkali metals. The PBEsol
 improves the performance and yields the least MAE for the alkali metals. Regarding the
 performance of the meta-GGAs, all functionals overestimate the equilibrium shortest
 distances except the M06L, TMTPSS and TM functionals. The M06L massively underestimates
 the equilibrium shortest distances of alkali metals, whereas, TMTPSS and TM agree very
 well with the experimental values. Both TMTPSS and TM functionals produce much better
 results than SCAN meta-GGA functional, though the SCAN contains intermediate
 van-der-Waals interactions.
 
  \textbf{Alkaline-earth metals :}~Unlike alkali metals, a reasonably good performance is
  observed using the PBE functionals for the alkaline-earth metals. However, the PBEsol
  underestimates the equilibrium shortest distances. Within meta-GGA functionals, TPSS and
  MS2 perform quite well. A sizable underestimation in the equilibrium shortest distances
  is observed using the M06L functional. MS0 and MS1 overestimate the equilibrium shortest
  distances, whereas, the overestimation percentage comparatively inadequate for the SCAN
  functional. In this case, both the TMTPSS and TM functionals underestimate the
  equilibrium shortest distances. But, the TMTPSS quite close to the experimental
  equilibrium shortest distances. In this case, the SCAN functional agrees well with the
  experimental values than TMTPSS and TM functionals.

 \textbf{Other transition metals :}~Due to the complicated structure of the Mn, La, and Hg,
 we separate out these elements from others. Regarding the performance of PBE, it massively
 overestimates the equilibrium shortest distances for all these metals. Reasonably good
 performance is observed using PBEsol. For meta-GGAs, all perform equivalently to predict
 the equilibrium shortest distances except M06L. M06L overestimates the shortest distances
 in a sizable order. The TPSS, MS0, MS1 perform equivalently. We obtain MAE $4.250$ pm from
 the revTPSS functionals. The MS2 and SCAN also produce the same amount of error in this
 case. Here, we obtain least MAE with the MS0 and MS1 functionals. TM functional 
 is the second best after MS0 and MS1 with MAE $2.7$ pm. In this case also the performance 
 TM functional is better to compare to the SCAN functional.

 \textbf{Overall performance :}~Correspond to the overall ranking we obtain the best MAE
 with the TM functional. Next, the performance of the TMTPSS is found to be best. The
 performance of TM and TMTPSS is quite well compared to the SCAN functional. The revTPSS,
 MS2 and SCAN functionals perform equivalently. The MAE of TPSS is less compare to the MS0
 and MS1. The M06L gives the largest MAE of order $8.1$ pm.

\subsection{Bulk Moduli}
 
The bulk modulus ($B_0$) is defined as the variation of the volume ($V$) due to the
external pressure ($P$). In DFT the bulk modulus is measured at the equilibrium
lattice constant $a_0$ or volume ($V_0$) as,

   \begin{longtable*}{lcccccccccccccccc}
\label{latt-cons}\\
 \caption{Bulk moduli ($B_0$) (in GPa) calculated using different solid structures using
 PBE, PBEsol, TPSS, revTPSS, M06L, MS0, MS1, MS2, SCAN, TMTPSS and TM functionals. The
 experimental values corrected for the finite thermal corrections. All the corrected values
 are taken from references~\cite{PJanthon14,GCsonks09}. For elements Mn, La and Hg the
 finite temperature corrected values are not available and we reported only the
 experimental values taken from reference~\cite{PJanthon13}. The the total mean 
 error (TME) are reported without  considering the Mn, La and Hg results and 
 with the values of Mn, La and Hg results. }\\
\hline\hline
Metals&PBE&PBEsol&TPSS&revTPSS&M06L&MS0&MS1&MS2&SCAN&TMTPSS&TM&Expt.\\
\hline\hline 
Sc	&	52.6	&	55.4	&	54.6	&	55.6	&	62.6	&	54.2	&	53.6	&	56.0	&	59.8	&	61.8	&	57.4	&	55.6	\\
Ti	&	116.8	&	125.6	&	123.2	&	125.8	&	128.6	&	124.0	&	122.4	&	127.6	&	125.2	&	125.8	&	127.0   &	108.3	\\
V	&	187.8	&	204.0	&	201.8	&	205.6	&	198.0	&	200.8	&	199.6	&	207.0	&	203.8	&	201.8	&	181.8	&	158.9	\\
Cr	&	263.6	&	288.2	&	283.2	&	291.8	&	276.8	&	284.8	&	281.2	&	292.0	&	280.2	&	283.6	&	286.8	&	174.5	\\
Fe	&	166.6	&	208.6	&	310.4	&	320.6	&	300.0	&	316.2	&	217.8	&	294.2	&	316.6	&	312.6	&	316.0	&	169.8	\\
Co	&	212.6	&	285.8	&	280.2	&	291.4	&	263.6	&	294.2	&	288.0	&	268.8	&	297.6	&	244.2	&	289.6	&	193.0	\\
Ni	&	197.7	&	230.1	&	226.2	&	241.8	&	221.9	&	242.2	&	235.7	&	244.8	&	238.3	&	233.1	&	236.6	&	185.5	\\
Cu	&	137.1	&	163.3	&	156.5	&	170.5	&	151.9	&	158.9	&	146.9	&	155.9	&	152.4	&	161.2	&	164.2	&	140.3	\\
Zn	&	74.0	&	91.8	&	86.0	&	97.8	&	74.2	&	99.4	&	82.8	&	102.0	&	105.2	&	96.8	&	105.6	&	69.7	\\
	&		&		&		&		&		&		&		&		&		&		&		&		\\
ME	&	17.0	&	44.1	&	51.8	&	60.6	&	46.9	&	57.7	&	41.4	&	54.7	&	58.2	&	51.7	&	56.6	&		\\
MAE	&	19.1	&	44.2	&	52.1	&	60.6	&	46.9	&	58.0	&	41.8	&	54.7	&	58.2	&	51.7	&	56.6	&		\\
	&		&		&		&		&		&		&		&		&		&		&		&		\\
Y	&	39.6	&	42.0	&	40.2	&	40.2	&	44.4	&	37.4	&	37.2	&	36.6	&	36.4	&	38.4	&	41.4	&	41.7	\\
Zr	&	92.8	&	98.8	&	96.4	&	97.0	&	95.4	&	92.8	&	92.2	&	95.6	&	95.8	&	96.2	&	97.0	&	95.9	\\
Nb	&	172.0	&	186.8	&	183.4	&	187.6	&	169.0	&	181.6	&	180.4	&	183.4	&	180.8	&	181.2	&	183.2	&	172.0	\\
Mo	&	266.4	&	289.0	&	278.6	&	286.2	&	260.2	&	287.0	&	285.2	&	287.4	&	284.0	&	278.6	&	283.6	&	264.7	\\
Tc	&	301.4	&	330.6	&	317.0	&	329.0	&	292.2	&	334.6	&	331.2	&	331.2	&	324.2	&	319.0	&	325.2	&	303.1	\\
Ru	&	316.4	&	353.8	&	334.4	&	350.4	&	302.0	&	365.4	&	361.4	&	356.8	&	342.2	&	337.0	&	345.8	&	317.7	\\
Rh	&	254.8	&	295.8	&	276.7	&	292.4	&	235.0	&	298.4	&	294.7	&	292.9	&	290.3	&	275.9	&	284.0	&	288.7	\\
Pd	&	165.3	&	201.8	&	187.9	&	201.1	&	148.3	&	199.5	&	195.3	&	200.1	&	193.5	&	189.5	&	195.3	&	195.4	\\
Ag	&	86.1	&	112.3	&	102.3	&	113.0	&	89.1	&	103.7	&	100.9	&	110.7	&	105.4	&	109.3	&	113.0	&	103.8	\\
Cd	&	40.8	&	59.4	&	54.4	&	62.2	&	58.6	&	59.6	&	56.0	&	63.4	&	57.4	&	64.8	&	50.2	&	53.8	\\
	&		&		&		&		&		&		&		&		&		&		&		&		\\
ME	&	-10.1	&	13.4	&	3.5	&	12.2	&	-14.3	&	12.3	&	9.8	&	12.1	&	7.3	&	5.3	&	8.2	&		\\
MAE	&	10.5	&	13.4	&	8.0	&	12.5	&	15.8	&	13.8	&	12.0	&	13.2	&	8.8	&	9.7	&	9.9	&		\\
	&		&		&		&		&		&		&		&		&		&		&		&		\\
Hf	&	110.0	&	116.6	&	114.2	&	116.4	&	114.2	&	113.8	&	112.8	&	117.0	&	116.6	&	117.4	&	118.2	&	109.7	\\
Ta	&	199.0	&	199.0	&	207.8	&	213.0	&	196.8	&	210.4	&	208.8	&	213.4	&	212.0	&	210.8	&	212.6	&	193.7	\\
W	&	309.6	&	309.6	&	324.6	&	336.0	&	310.6	&	338.8	&	332.2	&	336.0	&	331.4	&	330.2	&	334.4	&	312.3	\\
Re	&	370.6	&	399.6	&	390.8	&	406.8	&	392.8	&	417.8	&	410.2	&	408.4	&	412.6	&	398.8	&	404.8	&	368.8	\\
Os	&	405.6	&	443.6	&	426.8	&	450.8	&	409.0	&	466.6	&	461.0	&	455.6	&	459.6	&	440.6	&	449.4	&	424.6	\\
Ir	&	350.5	&	391.0	&	369.9	&	394.2	&	343.8	&	416.6	&	409.8	&	402.0	&	415.9	&	382.7	&	392.8	&	365.2	\\
Pt	&	245.0	&	285.8	&	264.6	&	284.1	&	224.2	&	305.9	&	298.8	&	294.4	&	244.1	&	271.2	&	280.4	&	284.2	\\
Au	&	131.1	&	164.1	&	157.6	&	162.7	&	127.7	&	168.2	&	172.5	&	167.8	&	158.2	&	153.1	&	159.5	&	174.8	\\
	&		&		&		&		&		&		&		&		&		&		&		&		\\
ME	&	-14.0	&	9.5	&	2.9	&	16.3	&	-14.3	&	25.6	&	21.6	&	20.2	&	14.6	&	8.9	&	14.9	&		\\
MAE	&	15.8	&	12.9	&	12.1	&	19.4	&	22.2	&	27.3	&	22.2	&	21.9	&	28.8	&	17.6	&	19.6	&		\\
	&		&		&		&		&		&		&		&		&		&		&		&		\\
K	&	3.5	&	3.5	&	3.3	&	3.3	&	3.2	&	3.1	&	3.0	&	3.2	&	3.2	&	3.8	&	3.8	&	3.7	\\
Rb	&	2.8	&	2.9	&	3.7	&	3.5	&	4.7	&	3.6	&	2.8	&	3.3	&	3.1	&	3.1	&	3.1	&	2.9	\\
Cs	&	2.0	&	2.0	&	1.8	&	1.9	&	4.1	&	2.0	&	2.0	&	2.0	&	2.0	&	2.1	&	2.1	&	2.1	\\
	&		&		&		&		&		&		&		&		&		&		&		&		\\
ME	&	-0.1	&	-0.1	&	0.0	&	0.0	&	1.1	&	0.0	&	-0.3	&	-0.1	&	-0.1	&	0.1	&	0.1	&		\\
MAE	&	0.1	&	0.1	&	0.5	&	0.4	&	1.4	&	0.5	&	0.3	&	0.3	&	0.3	&	0.1	&	0.1	&		\\
	&		&		&		&		&		&		&		&		&		&		&		&		\\
Ca	&	16.8	&	17.2	&	16.8	&	17.1	&	20.9	&	18.2	&	17.5	&	17.7	&	17.6	&	18.4	&	18.5	&	18.4	\\
Sr	&	11.5	&	12.3	&	11.4	&	11.7	&	17.1	&	11.8	&	11.4	&	11.9	&	11.1	&	12.5	&	12.6	&	12.4	\\
Ba	&	8.7     &	9.3 	&	8.4	    &	8.7	    &	11.8	&	7.7	    &	7.6	    &	8.3	    &	8.1	    &	9.2	    &	9.2	    &	9.3 	\\
	&		&		&		&		&		&		&		&		&		&		&		&		\\
ME	&	-1.0	&	-0.4	&	-1.2	&	-0.9	&	3.2	&	-0.8	&	-1.2	&	-0.7	&	-1.1	&	0.0	&	0.1	&		\\
MAE	&	1.0	&	0.4	&	1.2	&	0.9	&	3.2	&	0.8	&	1.2	&	0.7	&	1.1	&	0.1	&	0.1	&		\\
	&		&		&		&		&		&		&		&		&		&		&		&		\\
	\hline\hline
TME	&	-1.9	&	18.3	&	15.8	&	24.1	&	5.4	&	25.6	&	19.3	&	23.4	&	21.5	&	17.9	&	21.5	&		\\
TMAE	&	12.3	&	19.3	&	19.7	&	25.1	&	23.4	&	26.7	&	20.6	&	24.3	&	25.6	&	21.3	&	23.2	&		\\
\hline\hline
Mn	&	183.2	&	305.4	&	299.4	&	309.1	&	298.0	&	311.0	&	306.6	&	312.6	&	315.3	&	302.6	&	306.5	&	90.4	\\
La	&	24.2	&	26.4	&	25.8	&	25.4	&	28.0	&	22.8	&	22.6	&	23.6	&	24.2	&	25.6	&	24.2	&	26.6	\\
Hg	&	9.6  	&	36.0	&	22.4	&	32.4	&	19.6	&	24.4	&	21.6	&	24.0	&	21.2	&	33.6	&	38.8	&	28.2	\\
	&		&		&		&		&		&		&		&		&		&		&		&		\\
ME	&	23.9	&	74.2	&	67.5	&	73.9	&	66.8	&	71.0	&	68.5	&	71.7	&	71.8	&	72.2	&	74.8	&		\\
MAE	&	37.9	&	74.3	&	71.9	&	74.7	&	72.5	&	76.1	&	75.6	&	76.5	&	78.1	&	72.9	&	76.4	&		\\
	&		&		&		&		&		&		&		&		&		&		&		&		\\

\hline\hline
TME	&	0.2	&	23.0	&	20.1	&	28.3	&	10.5	&	29.4	&	23.4	&	27.4	&	25.7	&	22.4	&	26.0	&		\\
TMAE	&	14.5	&	23.8	&	24.0	&	29.3	&	27.5	&	30.8	&	25.1	&	28.7	&	30.0	&	25.6	&	27.7	&		\\

\hline
\hline

\end{longtable*}

\begin{figure*}
\includegraphics[width=7.0in,height=4.0in,angle=0.0]{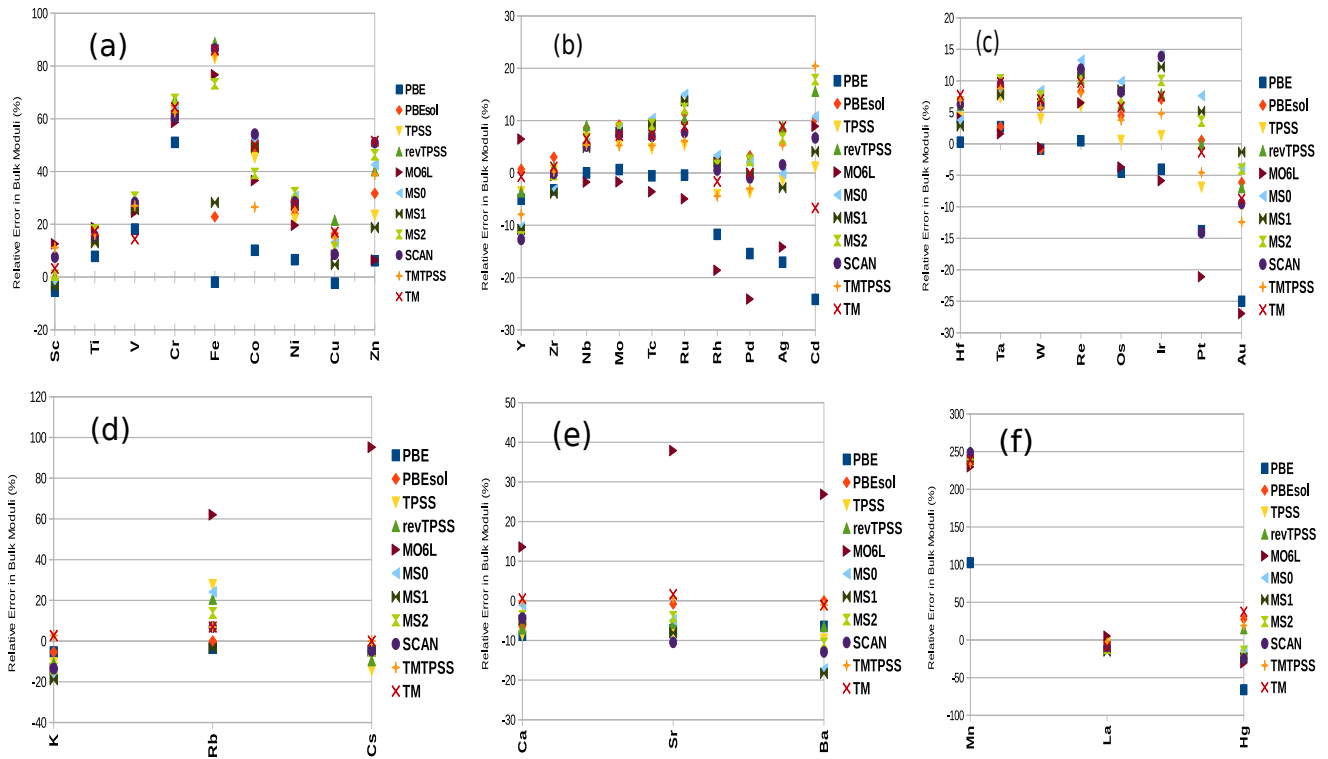}
\caption{Histograms of relative error in bulk moduli (in \%) are presented. The numbering
of the figures are as the order of the solids presented in TABLE II.}
\label{fig-bulk-moduli}
\end{figure*}

\begin{equation}
    B_0 = -V_0\Big(\frac{\partial P}{\partial V}\Big)_{a=a_0}~.
\end{equation}
Several equations of state (EOS)~\cite{EOS1,EOS2,EOS3,EOS4} are available to fit the energy
versus volume curve to obtain the bulk moduli. However, in the present case, we use the
Birch-Murnaghan equation of state to fit and obtain the bulk moduli of alkali,
alkaline-earth, and transition metals. It is well known that determining the bulk modulus
posses a great challenge, in particular for the transition metals~\cite{PJanthon14}. The
experimental values along with all the functionals values are presented in TABLE II. The
general trend of the arrangement of TABLE II is the same as it is done in the case of the
equilibrium shortest distances. The $3d$, $4d$ and $5d$ band elements are separated out.
The values of alkali metals and alkali earth metals are also shown separately. The Mn, La,
and Hg values are also separated out. In Fig. (\ref{fig-bulk-moduli}) we also plot the
percentage deviation of all the metals considered in our work. The Fig.
(\ref{fig-bulk-moduli}) is also arranged according to the data presented in TABLE II.  

\textbf{$3d$ transition metals :}~The PBE functional works well for the $3d$ transition
metals and produces the least MAE. PBE functional overestimates the bulk moduli for few
cases, whereas, underestimates in results are observed for others. However, the performance
of the revised version of PBE i.e., PBEsol deviates more from the experimental values and
yield  MAE $44.178$ GPa. Regarding the meta-GGAs, all functionals deviate from the
experimental values more or less. However, MS1 reports being least MAE among the meta-GGA
functionals. We obtain equivalent performance using the TPSS, revTPSS, M06L, MS0, MS2, SCAN,
TMTPSS and TM functionals. 

\textbf{$4d$ transition metals :}~For the $4d$ transition metals the TPSS functional
performs best with MAE $8.0$ GPa. The PBE functional underestimates the bulk moduli of all
the elements except Nb, Mo Tc, Ru. For those metals, the PBE values are very close to the
experimental one. Unlike PBE, the PBEsol overestimates the bulk moduli of all the $4d$
metals and provides the MAE $13.4$ GPa. The performance of revTPSS, MS0, MS1, MS2 are
almost equivalent. All functionals overestimate the bulk moduli except few cases. The
``MS'' functionals yield reasonably good bulk moduli for Y, Zr, Pd, Ag, and Cd. We find
reasonably good performance for Y, Zr, and Ag using revTPSS functional. For others
overestimation is observed. The bulk moduli obtained from SCAN functional matches well with
the experimental values except for few cases like Mo, Tc, and Ru. For Mo, Tc, and Ru 
the SCAN functional overestimates the value. From the TMTPSS and TM functionals, we obtain the
similar trend as it is observed for SCAN. In this case, the TM functional overestimates
more than TMTPSS in predicting the bulk moduli of $4d$ metals.

\textbf{$5d$ transition metals :}~For the $5d$ transition metals the performance of
revTPSS, TPSS, and PBE are found to be better compared to the others. The M06L
underestimates the bulk moduli for all, except Hf, Ta, and Re. The overall ME obtain from
the M06L is found to be negative. The MS0, MS1, MS2, and SCAN functionals perform
equivalently and provide almost equivalent ME and MAE. The performance of TM functional
indicates that it yields slightly greater values for bulk moduli compared to the TMTPSS
functional. Comparing the performance of SCAN, TMTPSS and TM functionals, both the TMTPSS 
and TM show improve performance than SCAN functional.

\textbf{Alkali metals :}~The alkali matters are considered as a ``soft-matter'' due to the
smaller extent of their bulk moduli. The bulk moduli of alkali metals vary only from $2.1$
GPa (Cs) to $3.7$ GPa (K). All the functionals perform reasonably good in case of all the
alkali metals. The largest MAE is obtained from M06L functional, whereas, all other
meta-GGA and GGAs varies from $0.1$ GPa to  $0.5$ GPa.

\textbf{Alkaline-earth metals :}~Like the alkali metals, we obtain equivalent performance
of all the functionals for alkaline-earth metals. In this case, the largest MAE is obtained
from M06L. The performance of revTPSS, MS0, MS1, MS2, and SCAN are almost equivalent. The
TPSS functional yields the MAE $0.4$ GPa. In this case, we obtain the least MAE from the
TMTPSS and TM functional. Both functionals perform fairly well than SCAN functional. 

 \textbf{Other transition metals :}~The bulk moduli of Mn, La, and Hg are problematic within (most of) 
 the density functionals. The experimental bulk modulus of Mn is $90.4$
 GaP, whereas, all the functionals show sizable overestimation. The bulk modulus of La
 using all the functionals are predicted to be fairly well, but overestimation and
 underestimation in values are observed for Hg using all the functionals. The bulk modulus
 of Hg obtained from TPSS, revTPSS, MS0, MS1, MS2, SCAN and TMTPSS show good agreement with
 the experimental values. The M06L underestimates the bulk modulus of Hg and the same
 amount of overestimation is observed for the TM functional.    
 
 \textbf{Overall performances :}~To evaluate the overall performance of all the functionals
 we present the ME and MAE with and without considering Mn, La and Hg transition metals.
 Overall, the PBE functional yields the best MAE ($14.5$ GPa with other transition metals
 and $12.3$ without other transition metals). All meta-GGAs overall overestimate the ME and
 MAE. The performance of TPSS functional seems to be the best within all the meta-GGAs.

\subsection{Cohesive Energies}

The Cohesive energies are equivalent to the atomization energies in the case of the bulk 
solids. It is expressed as the energy per atom as,

\begin{longtable*}{lcccccccccccccccc}
\label{latt-cons}\\
 \caption{Fixed lattice constant cohesive energies (in eV/atom) of
different solid structures using PBE, PBEsol, TPSS, revTPSS, M06L, MS0, MS1, MS2, SCAN, TMTPSS and TM functionals. All the finite temperature corrected experimental values are collected from  references~\cite{PJanthon14,GKresse13}. For elements Mn, La and Hg the temperature corrected values are not available and we report only the experimental values taken from reference~\cite{PJanthon13}.}\\
\hline\hline
Metals&PBE&PBEsol&TPSS&revTPSS&M06L&MS0&MS1&MS2&SCAN&TMTPSS&TM&Expt.\\
\hline\hline 
Sc	&	4.20	&	4.58	&	4.47	&	4.58	&	5.08	&	4.39	&	4.31	&	4.49	&	4.37	&	4.69	&	4.81	&	3.93	\\
Ti	&	5.40	&	5.87	&	5.56	&	5.77	&	6.22	&	5.29	&	5.23	&	5.48	&	5.30	    &	6.01	&	6.04	&	4.88	\\
V	&	5.25	&	5.83	&	5.51	&	5.76	&	6.28	&	5.09	&	5.01	&	5.43	&	4.96	&	5.89	&	5.93	&	5.34	\\
Cr	&	4.05	&	4.71	&	4.24	&	4.47	&	4.50	    &	3.49	&	3.44	&	3.96	&	3.26	&	4.62	&	4.66	&	4.15	\\
Fe	&	4.89	&	5.61	&	5.32	&	5.59	&	5.03	&	5.16	&	5.08	&	5.51	&	5.13	&	5.68	&	5.80  	&	4.32	\\
Co	&	5.07	&	5.83	&	5.72	&	6.04	&	5.71	&	5.76	&	5.67	&	6.01	&	5.86	&	6.19	&	6.30	    &	4.47	\\
Ni	&	4.68	&	5.33	&	5.06	&	5.43	&	4.54	&	5.22	&	5.11	&	3.23	&	5.25	&	5.62	&	5.71	&	4.48	\\
Cu	&	3.48	&	4.03	&	3.75	&	4.09	&	3.06	&	3.80	&	3.71	&	4.09	&	3.87	&	4.32	&	4.38    &	3.51	\\
Zn	&	1.10	&	1.57	&	1.34	&	1.61	&	1.54	&	1.55	&	1.46	&	1.74	&	1.52	&	1.71	&	1.89	&	1.38	\\
	&		&		&		&		&		&		&		&		&		&		&		&		\\
ME	&	0.18	&	0.77	&	0.50	&	0.76	&	0.61	&	0.37	&	0.28	&	0.39	&	0.34	&	0.92	&	1.01	&		\\
MAE	&	0.30	&	0.77	&	0.51	&	0.76	&	0.71	&	0.57	&	0.52	&	0.71	&	0.62	&	0.92	&	1.01	&		\\
	&		&		&		&		&		&		&		&		&		&		&		&		\\
Y	&	4.21	&	4.60	&	4.43	&	4.57	&	5.06	&	4.36	&	4.30	&	4.49	&	4.42	&	4.70	&	4.81	&	4.42	\\
Zr	&	6.27	&	6.78	&	6.35	&	6.55	&	6.80	&	5.83	&	5.83	&	6.04	&	6.12	&	6.82	&	6.84	&	6.32	\\
Nb	&	6.79	&	7.47	&	7.14	&	7.40	&	8.63	&	6.85	&	6.75	&	7.07	&	6.56	&	7.45	&	7.49	&	7.47	\\
Mo	&	6.35	&	7.18	&	6.63	&	6.95	&	6.85	&	6.34	&	6.23	&	6.61	&	5.81	&	6.96	&	7.03	&	6.84	\\
Tc	&	6.90	&	7.85	&	7.17	&	7.59	&	6.53	&	7.28	&	7.16	&	7.57	&	6.72	&	7.61	&	7.78	&	7.17	\\
Ru	&	6.88	&	7.87	&	7.21	&	7.66	&	6.25	&	7.69	&	7.54	&	7.79	&	7.56	&	7.70	&	7.85	&	6.80	\\
Rh	&	5.70	&	6.66	&	6.00	&	6.40	&	5.40	&	5.9	&	5.81	&	6.18	&	5.58	&	6.41	&	6.55	&	5.76	\\
Pd	&	3.74	&	4.47	&	4.00	&	4.39	&	4.17	&	4.24	&	4.13	&	4.46	&	4.38	&	4.61	&	4.71	&	3.93	\\
Ag	&	2.52	&	3.08	&	2.73	&	3.03	&	3.24	&	2.79	&	2.70	&	3.10	&	2.88	&	3.29	&	3.35	&	2.96	\\
Cd	&	0.73	&	1.16	&	0.95	&	1.2	&	1.33	&	1.08	&	1.01	&	1.34	&	1.03	&	1.39	&	1.49	&	1.18	\\
	&		&		&		&		&		&		&		&		&		&		&		&		\\
ME	&	-0.28	&	0.43	&	-0.02	&	0.29	&	0.14	&	-0.05	&	-0.14	&	0.18	&	-0.18	&	0.41	&	0.51	&		\\
MAE	&	0.29	&	0.43	&	0.18	&	0.30	&	0.45	&	0.34	&	0.34	&	0.36	&	0.42	&	0.41	&	0.51	&		\\
	&		&		&		&		&		&		&		&		&		&		&		&		\\
Hf	&	6.48	&	7.14	&	6.77	&	7.08	&	7.40	&	6.82	&	6.73	&	7.03	&	6.35	&	7.05	&	7.22	&	6.44	\\
Ta	&	8.58	&	9.23	&	8.67	&	9.03	&	9.00	&	8.99	&	8.85	&	9.20	&	8.90	&	9.53	&	9.55	&	8.11	\\
W	&	8.47	&	9.27	&	8.83	&	9.22	&	9.77	&	9.31	&	9.12	&	9.46	&	9.09	&	9.41	&	9.45	&	8.83	\\
Re	&	7.79	&	8.77	&	8.22	&	8.65	&	7.74	&	8.89	&	8.68	&	8.95	&	8.51	&	8.71	&	8.91	&	8.06	\\
Os	&	8.30	&	9.36	&	8.80	&	9.10	&	8.23	&	9.53	&	9.34	&	9.53	&	9.08	&	9.26	&	9.49	&	8.22	\\
Ir	&	7.19	&	8.27	&	7.56	&	8.10	&	6.97	&	8.51	&	8.31	&	8.43	&	8.37	&	8.07	&	8.27	&	6.96	\\
Pt	&	5.42	&	6.27	&	5.74	&	6.18	&	5.81	&	6.31	&	6.15	&	6.42	&	6.17	&	6.31	&	6.46	&	5.87	\\
Au	&	3.03	&	3.72	&	3.27	&	3.60	&	3.57	&	3.59	&	3.45	&	3.81	&	3.55	&	3.83	&	3.93	&	3.83	\\
	&		&		&		&		&		&		&		&		&		&		&		&		\\
ME	&	-0.13	&	0.71	&	0.19	&	0.58	&	0.27	&	0.70	&	0.54	&	0.81	&	0.46	&	0.73	&	0.87	&		\\
MAE	&	0.34	&	0.74	&	0.37	&	0.64	&	0.43	&	0.76	&	0.63	&	0.82	&	0.56	&	0.73	&	0.87	&		\\
	&		&		&		&		&		&		&		&		&		&		&		&		\\
K	&	0.86	&	0.92	&	0.92	&	0.94	&	0.66	&	0.86	&	0.84	&	0.88	&	0.84	&	0.99	&	0.99	&	0.94	\\
Rb	&	0.78	&	0.84	&	0.81	&	0.83	&	1.30	&	0.78	&	0.76	&	0.80	&	0.76	&	0.93	&	0.92	&	0.86	\\
Cs	&	0.72	&	0.78	&	0.74	&	0.77	&	1.35	&	0.73	&	0.71	&	0.76	&	0.70	&	0.88	&	0.88	&	0.81	\\
	&		&		&		&		&		&		&		&		&		&		&		&		\\
ME	&	-0.08	&	-0.02	&	-0.05	&	-0.02	&	0.23	&	-0.08	&	-0.10	&	-0.06	&	-0.10	&	0.06	&	0.06	&		\\
MAE	&	0.08	&	0.02	&	0.05	&	0.02	&	0.42	&	0.08	&	0.10	&	0.06	&	0.10	&	0.06	&	0.06	&		\\
	&		&		&		&		&		&		&		&		&		&		&		&		\\
Ca	&	1.91	&	2.11	&	2.02	&	2.08	&	2.50	&	2.00	&	1.96	&	2.03	&	2.08	&	2.17	&	2.29	&	1.86	\\
Sr	&	1.61	&	1.81	&	1.76	&	1.83	&	2.27	&	1.79	&	1.74	&	1.84	&	1.82	&	1.96	&	2.06	&	1.73	\\
Ba	&	1.88	&	2.12	&	2.03	&	2.11	&	2.51	&	2.00	&	1.95	&	2.1	&	2.04	&	2.24	&	2.34	&	1.91	\\
	&		&		&		&		&		&		&		&		&		&		&		&		\\
ME	&	-0.03	&	0.18	&	0.10	&	0.17	&	0.59	&	0.10	&	0.05	&	0.16	&	0.15	&	0.29	&	0.40	&		\\
MAE	&	0.07	&	0.18	&	0.10	&	0.17	&	0.59	&	0.10	&	0.05	&	0.16	&	0.15	&	0.29	&	0.40	&		\\
	&		&		&		&		&		&		&		&		&		&		&		&		\\
Mn	&	3.80	&	4.55	&	3.95	&	4.06	&	2.81	&	3.04	&	3.01	&	3.54	&	2.90	&	4.13	&	4.28	&	2.92	\\
La	&	4.30	&	4.79	&	4.51	&	4.56	&	5.03	&	3.83	&	3.82	&	4.06	&	3.72	&	4.58	&	4.61	&	4.47	\\
Hg	&	0.15	&	0.54	&	0.22	&	0.43	&	0.55	&	0.43	&	0.32	&	0.64	&	0.39	&	0.67	&	0.75	&	0.62	\\
	&		&		&		&		&		&		&		&		&		&		&		&		\\
ME	&	0.08	&	0.62	&	0.22	&	0.35	&	0.13	&	-0.24	&	-0.29	&	0.08	&	-0.33	&	0.46	&	0.54	&		\\
MAE	&	0.51	&	0.68	&	0.49	&	0.47	&	0.25	&	0.32	&	0.35	&	0.35	&	0.33	&	0.46	&	0.54	&		\\
	&		&		&		&		&		&		&		&		&		&		&		&		\\

	\hline\hline

TME	&	-0.06	&	0.53	&	0.19	&	0.44	&	0.33	&	0.22	&	0.12	&	0.34	&	0.11	&	0.57	&	0.67	&		\\
TMAE	&	0.29	&	0.55	&	0.31	&	0.47	&	0.50	&	0.45	&	0.41	&	0.51	&	0.44	&	0.57	&	0.67	&		\\

\hline
\hline

\end{longtable*}

 \begin{figure*}
\includegraphics[width=7.0in,height=4.0in,angle=0.0]{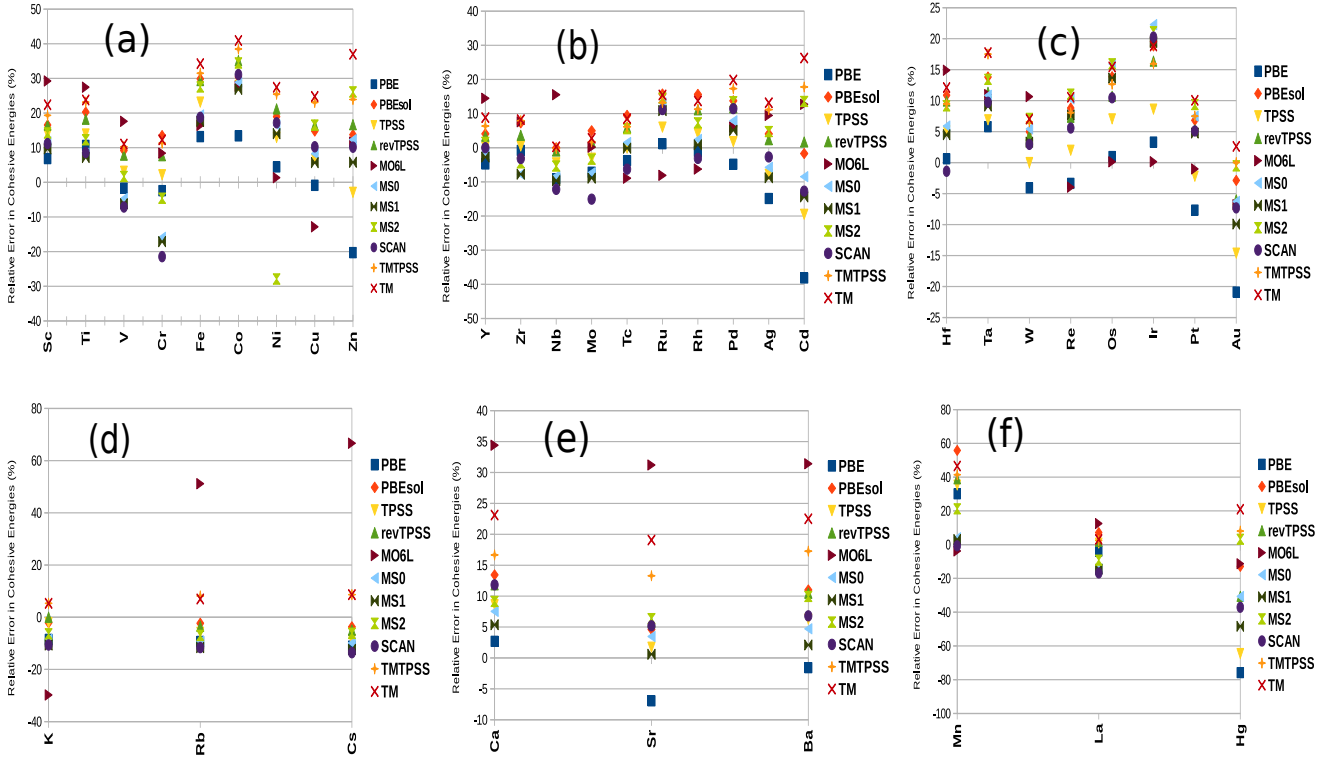}
\caption{Histograms of relative error in cohesive energies (in \%) are presented. The
numbering of the figures are as the order of the solids presented in TABLE III.}
\label{fig-cohesive-energy}
\end{figure*}

 \begin{equation}
 E_{coh}= E_{atom} - \frac{E_{bulk}}{N}~,
 \end{equation}
 where $E_{atom}$ is the atomic energy and $E_{bulk}$ is the bulk energy of unit cell
 having $N$ atoms. Predicting the cohesive energies of transition metals posses great
 challenge because of their ``strongly'' correlated nature~\cite{GKresse13}. TABLE III
 presents the performance of all the functionals along with the experimental values. We
 arrange TABLE III in the same manner as it is done in the case of lattice constants and
 bulk moduli. The behavior of all the functionals are also plotted in the
 Fig.~\ref{fig-cohesive-energy}. The percentage deviation of all the functionals is shown
 there. All the cohesive energies are calculated at the equilibrium lattice constants of
 the functional.  
 
 \textbf{$3d$ transition metals :}~The PBE functional predicts the cohesive energies of
 $3d$ transition metals quite well compared to the other GGA and meta-GGA based
 functionals. Though PBE overestimates the cohesive energies for Sc, Ti, Fe, and Co, but
 overall both the ME and MAE are found to be reasonably well predicted within PBE
 functional. The PBEsol overestimates the cohesive energies of $3d$ transition metals
 more. Therefore, increase in ME and MAE is observed with the PBEsol functional. Within
 the meta-GGA functionals, the performance of MS1 is the best. MS0 quite closely follows
 the MS1 functional. Comparing the SCAN, TMTPSS and TM functionals, the SCAN functional
 performs better compared to TM based functionals. The revTPSS overestimates the TPSS
 values more and yield more enhanced ME and MAE. The M06L yield the same MAE as it is
 obtained from MS2.
 
 A noticeable observation is that revTPSS behaves closely as PBEsol. This can be
 understood from the explanation given in ref.~\cite{GKresse13}. In metals, as the  $d$
 band started filling, the meta-GGA total charge density becomes the sum of several
 one-electron orbitals~\cite{GKresse13}. Therefore, revTPSS becomes PBEsol like. But, the
 improved functionals like SCAN behaves more closely to PBE for Sc, Ti. But the SCAN
 functional reasonably underestimates the cohesive energies of V, and Cr, and
 overestimates for Fe to Zn. A different tendency is observed for TMTPSS and TM
 functionals. Both overestimates the cohesive energies noticeably.     
 
 \textbf{$4d$ transition metals :}~Unlike previous observation, the PBE functional
 underestimates the cohesive energies of $4d$ transition metals. The PBEsol overestimates
 the cohesive energies of all metals except Nb. In this case, the TPSS predicts the best
 ME and MAE. The TPSS functional predicts reasonably good cohesive energies for Y, Zr, Tc,
 and Pd. For others, it underestimates or overestimates the values. In this case, also we
 observe that the revTPSS follows closely that of the PBEsol results. The M06L functional
 overall overestimates the cohesive energies except for few cases. The MS0, MS1 and MS2
 functionals perform almost equivalently. The SCAN functional are also closely following
 the ``MS'' functionals. We observe that the TMTPSS and TM functionals show PBEsol like
 tendency. Here, the TM functional overestimates the cohesive energies more which are
 produced by TMTPSS.

 \textbf{$5d$ transition metals :}~For $5d$ transition metals also, the PBE performs quite
 reasonably. PBE overestimates the cohesive energies for Ta, and Ir but performs well for
 Hf, W, Re, Os, Pt, and Au. The PBEsol shows the same tendency as it is obtained in the
 cases of $3d$ and $4d$ elements. The TPSS functional enhances the PBE results, therefore,
 the increase in the MAE is observed compared to the PBE results. Here, the revTPSS also
 follows closely the PBEsol results. The M06L functionals overestimate the half $d$ fill
 transition metals but perform well for largely filled $d$ transition metals. In this case
 also the ``MS'' functionals perform equivalently, though the results of MS2 are found to
 be more enhanced than MS0 and MS1. In this case, the SCAN functional performs quite well
 compared to the TMTPSS and TM functionals. Both the TMTPSS and TM functional overestimate
 the results of all the $5d$ metals, but the overestimation tendency is lesser than the
 SCAN functional.

\textbf{Alkali metals :}~The alkali metals are often included in the benchmarking
calculations for the different functionals for the semiconductor in predicting the
cohesive energies. All the GGAs and meta-GGAs (except M06L) perform quite reasonably for
the cohesive energies of alkali metals. The PBEsol and revTPSS show the best performance
with MAE $0.02$ eV/atom. In this case, the ``MS'' functionals and SCAN perform
equivalently. The TMTPSS and TM also predict the cohesive energies which are quite close
to the experimental results.

\textbf{Alkaline-earth metals :}~For alkaline-earth metals, PBE and TPSS perform quite
remarkably. The PBE and TPSS show opposite tendency. PBE shows a small degree of
underestimation (except Ca), whereas, TPSS shows a small degree of overestimation. The
same behavior is observed for the case of PBEsol and revTPSS. In this case, also the
``MS'' functionals and SCAN perform equivalently. We observe a noticeable amount of
overestimation in the performance of TMTPSS and TM functionals. The performance of SCAN
functional is quite better compared to the TM based functionals.

 \textbf{Other transition metals :}~The cohesive energy of Mn is largely overestimated
 within all the GGA and meta-GGA functionals except MS0, MS1 and SCAN functionals. Overall
 the M06L, ``MS'' and SCAN functionals predict the best MAE for all these metals. The TM
 functional predicts very well cohesive energies for La and Hg. As usual the TM functional
 enhance the error of TMTPSS more. Unlike other transition metals, in this case, we
 observe that the revTPSS and PBEsol results deviate from each other.     
 
 \textbf{Overall performances :}~Overall, the PBE performs quite reasonably for the
 cohesive energies. Within the meta-GGA functionals, the TPSS performs quite well and
 predicts the overall best MAE within the meta-GGAs. The PBEsol and revTPSS perform
 equivalently. We also obtain the same degree of overall ME and MAE using the M06L, MS0,
 MS1, MS2 and SCAN functionals. The TM functional deviates more from accuracy in
 predicting the cohesive energies of alkali, alkaline-earth, and transition metals. The
 errors obtained from the TM functional are often more enhanced than that of the TMTPSS functional.

  \begin{figure*}
\includegraphics[width=7.0in,height=3.5in,angle=0.0]{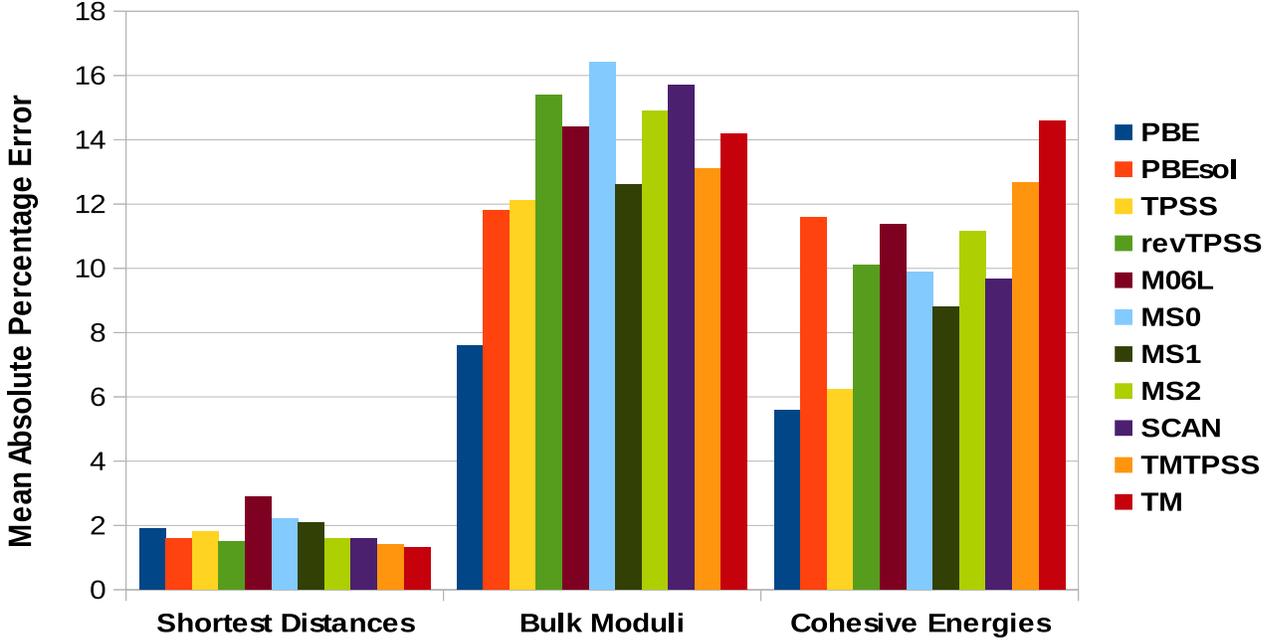}
\caption{Histograms of mean absolute percentage error (MAPE) in the shortest distances, bulk moduli and 
cohesive energies are presented. In the MAPE we excluded the Mn, La and Hg.}
\label{mape}
\end{figure*}

\begin{table*}
\label{stat-table}
 \caption{Statistical analysis and ranking of different functionals in the Shortest Distances, Cohesive Energies and Bulk Moduli.}
\begin{tabular}{c|  c  c  c  c  c   c   c  c c c c c c c}
\hline\hline
&&PBE&PBEsol&TPSS&revTPSS&M06L&MS0&MS1&MS2&SCAN&TMTPSS&TM\\
\hline\hline 
 	&	TME	&	3.8	&	-0.6	&	3.1	&	0.8	&	-1.7	&	2.6	&	2.7	&	1.2	&	1.2	&	1.1	&	-0.1	\\
Shortest	&	TMAE	&	5.4	&	4.4	&	5.1	&	4.3	&	8.1	&	6.1	&	5.9	&	4.6	&	4.6	&	3.9	&	3.7	\\
Distances	&	TMAPE	&	1.9	&	1.6	&	1.8	&	1.5	&	2.9	&	2.2	&	2.1	&	1.6	&	1.6	&	1.4	&	1.3	\\
	&	rank	&	8	&	4	&	7	&	3	&	11	&	10	&	9	&	4	&	4	&	2	&	1	\\
	&		&		&		&		&		&		&		&		&		&		&		&		\\
	\hline
	&	TME	&	-1.9	&	18.3	&	15.8	&	24.1	&	5.4	&	25.6	&	19.3	&	23.4	&	21.5	&	17.9	&	21.5	\\
Bulk 	&	TMAE	&	12.3	&	19.3	&	19.7	&	25.1	&	23.4	&	26.7	&	20.6	&	24.3	&	25.6	&	21.3	&	23.2	\\
Moduli	&	TMAPE	&	7.6	&	11.8	&	12.1	&	15.4	&	14.4	&	16.4	&	12.6	&	14.9	&	15.7	&	13.1	&	14.2	\\
	&	rank	&	1	&	2	&	3	&	9	&	7	&	11	&	4	&	8	&	10	&	5	&	6	\\
	&		&		&		&		&		&		&		&		&		&		&		&		\\
	\hline
	&	TME	&	-0.08	&	0.53	&	0.18	&	0.45	&	0.35	&	0.26	&	0.16	&	0.37	&	0.15	&	0.58	&	0.68	\\
Cohesive 	&	TMAE	&	0.26	&	0.54	&	0.29	&	0.47	&	0.53	&	0.46	&	0.41	&	0.52	&	0.45	&	0.59	&	0.68	\\
Energies	&	TMAPE	&	5.58	&	11.59	&	6.22	&	10.09	&	11.37	&	9.87	&	8.8	&	11.16	&	9.66	&	12.66	&	14.59	\\
	&	rank	&	1	&	9	&	2	&	6	&	8	&	5	&	3	&	7	&	4	&	10	&	11	\\

	\hline\hline

		&	Average rank	&	3.3	&	5	&	4	&	6	&	8.7	&	8.7	&	5.3	&	6.3	&	6	&	5.7	&	6	\\
\hline
\hline
\end{tabular}

\end{table*}

\section{Challenges of advance meta-GGA functionals}

In TABLE IV we present the overall statistical analysis and ranking of each functional best on their performance.  In Fig.(\ref{mape}), we
present the mean absolute percentage of the individual functionals
excluding the Mn, La and Hg. Based on these analyses it is indicative that
simultaneously predicting both the structural and energetic properties of
the transition metals within GGAs and meta-GGAs functionals possess great
challenge. Meta-GGA functionals belong to the third rung of the Jacob's
ladder. Within the meta-GGAs, the SCAN and TM functionals considered as one
of the most advanced functionals but the performance of those functionals
do not show any improvement over the PBE functionals for the bulk moduli
and cohesive energies. For the $4d$ and $5d$ transition metals, alkali
metals and alkaline-earth metals the SCAN and TM functionals improve the
performance but for the $3d$ transition metals PBE performs better. The SCAN, M06L and TM functionals capture mid-range vdW interactions, which accounts improve performance for some systems but remain difficult for other systems. Especially, for alkali metals, the TMTPSS and TM functionals show improvement over PBE, M06L, and SCAN functionals. In the present study, we do not include any long-range corrected vdW interactions into the functionals form. The inclusion of the long-range corrected vdW interactions may improve the functional performance by including the long-range electron correlation effect.

\section{Conclusions}

In this paper, we assess the performance of the recent meta-GGA density functionals along
with the popularly used GGA based functionals for the lattice constants, bulk moduli and
cohesive energies of alkali, alkaline-earth, and transition metals. The present paper is
arranged by addressing the performance of the $3d$, $4d$, $5d$, alkali, alkaline-earth,
and other transition metals. Due to the complicated structure of the Mn, La, and Hg, we
discuss these three materials separately. Special attention has been paid to the
performance on the recently proposed SCAN, TMTPSS and TM functionals.
Based on these analysis, benchmark calculations  and the level of deficiencies 
of all the functionals one conclude that:

(i) For the equilibrium shortest distances of the $3d$ transition metals PBE results
reasonably good compared to all other GGAs and meta-GGAs. For $4d$ and $5d$ transition
metals, alkali, and alkaline-earth metals the PBE results are too large except few cases.
The PBEsol and revTPSS follow essentially the identical results. The largest error is
obtained from the M06L functional. The performance of MS0 and MS1 are identical, whereas,
MS2 and SCAN closely follow each other in most of the cases. We obtain the best
performance with the TM functional. The TM perform even better than SCAN in most of the
cases except alkaline-earth metals. We found the performance of TMTPSS is quite close that 
of the TM functional for $3d$ transition metals.

(ii) Regarding the bulk moduli, the PBE functional outperforms all other semilocal GGAs
and meta-GGAs functionals in every case. Overall PBE underestimates the bulk moduli of
$4d$ transition metals, $5d$ transition metals, alkali metals and alkaline-earth metals. 
For $3d$ metals the PBE seems to overestimate the results. The bulk moduli of Mn is a difficult case for all
the GGAs and meta-GGAs functionals. In this case, the revTPSS results also follows very closely 
the PBEsol results. The performance of all other meta-GGAs vary noticeably from the PBE
results. Concerning the performance of TMTPSS and TM functional performance, both perform
well then the SCAN functional.

(iii) In predicting the cohesive energies of all the metals, PBE outperforms all other
functionals. Only for the alkali metals, the PBE results deviate from the experimental
results. The revTPSS closely follow the PBEsol in all the cases. Overall consideration
shows that the TM predicts the largest mean absolute error. Though in the case of the
alkali metals, the TMTPSS and TM functionals perform quite well compared to the other cases, but
overall both the functionals overestimate the cohesive energies. Overall, comparison of SCAN, TMTPSS
and TM show that the SCAN functional performs better compare to the TMTPSS and TM functional 
in estimation the cohesive energies.

\section{Acknowledgement}

S. J. would like to acknowledge the financial support from the Department of Atomic Energy, 
Government of India. K.S. would like to acknowledge the financial support from the Department 
of Science and Technology, Government of India, during his summer intern in NISER.


\begin{thebibliography}{unsrt} 

\bibitem{KS65} W. Kohn and L. J. Sham, Phys. Rev. {\bf 140}, A1133 (1965).

\bibitem{lda} J. P. Perdew and A. Zunger.  Phys. Rev. B, {\bf 23} 5048, (1981).



\bibitem{PW86} J. P. Perdew and Y. Wang, Phys. Rev. B {\bf 33}, 8800 (1986).

\bibitem{B88} A. D. Becke, Phys. Rev. A {\bf 38}, 3098 (1988).

\bibitem{LYP88} C. Lee, W. Yang, and R. G. Parr, Phys. Rev. B {\bf 37}, 785 (1988).

\bibitem{PW91} J. P. Perdew, J. A. Chevary, S. H. Vosko, K. A. Jackson, M. R. Pederson, D. J. Singh, and C. Fiolhais, Phys. Rev. B {\bf 46}, 6671 (1992).

\bibitem{B3PW91} A. D. Becke, J. Chem. Phys. {\bf 104}, 1040 (1996).


\bibitem{PBE96} J. P. Perdew, K. Burke, and M. Ernzerhof, Phys. Rev. Lett. {\bf 77}, 3865 (1996).


\bibitem{AE05} R. Armiento and A. E. Mattsson, Phys. Rev. B {\bf 72}, 085108 (2005).


\bibitem{ZWu06} Z. Wu and R. E. Cohen, Phys. Rev. B {\bf 73}, 235116 (2006).

\bibitem{SOGGA} Y Zhao, and D. G. Truhlar, J. Chem. Phys. 128, 184109 (2008).


\bibitem{PBEsol} J. P. Perdew, A. Ruzsinszky, G. I. Csonka, O. A. Vydrov, G. E. Scuseria, L. A. Constantin, X. Zhou, and K. Burke, Phys. Rev. Lett. {\bf 100}, 136406 (2008).


\bibitem{con1} L. A. Constantin, J. P. Perdew, and J. M. Pitarke, Phys. Rev. B 79, 075126 (2009).

\bibitem{con2} E. Fabiano, L. A. Constantin, and F. Della Sala, Phys. Rev. B 82, 113104 (2010).

\bibitem{con3} E. Fabiano, L. A. Constantin, and F. Della Sala, J. Chem. Theory Comput., 7 (11), pp 3548–3559 (2011).

\bibitem{con4} L. A. Constantin, A. Terentjevs, F. Della Sala, P. Cortona, and E. Fabiano, Phys. Rev. B 93, 045126 (2016).


\bibitem{BR89} A. D. Becke and M. R. Roussel, Phys. Rev. A {\bf 39}, 3761 (1989).



\bibitem{VSXC98} T. V. Voorhis and G. E. Scuseria, J. Chem. Phys. {\bf 109}, 400 (1998).








\bibitem{MO6L} Y. Zhao and D. G. Truhlar, J. Chem. Phys. {\bf 125}, 194101 (2006).

\bibitem{TPSS03} J. Tao, J. P. Perdew, V. N. Staroverov, and G. E. Scuseria, Phys. Rev. Lett. {\bf 91}, 146401 (2003).

\bibitem{revTPSS} J. P. Perdew, A. Ruzsinszky, G. I. Csonka, L. A. Constantin, and J. Sun, Phys. Rev. Lett. {\bf 103}, 026403 (2009).








\bibitem{con5} L. A. Constantin, E. Fabiano, and F. Della Sala, J. Chem. Theory Comput., 9 (5), pp 2256–2263 (2013).


\bibitem{MS0} J. Sun, B. Xiao, and A. Ruzsinszky,  J. Chem. Phys. 137, 051101 (2012). 

\bibitem{MS1MS2} J. Sun, R. Haunschild, B. Xiao, I. W. Bulik, G. E. Scuseria, and J. P. Perdew,  J. Chem. Phys. 138, 044113 (2013).

\bibitem{MSh} J. Sun, J. P. Perdew, and A. Ruzsinszky, Proceedings of the National Academy of Sciences of the United States of America, 112 (3) 685-689 (2015)


\bibitem{SCAN15} J. Sun, A. Ruzsinszky, and J. P. Perdew, Phys. Rev. Lett. {\bf 115}, 036402 (2015).


\bibitem{Tao-Mo16} J. Tao and Y. Mo, Phys. Rev. Lett. {\bf 117}, 073001 (2016).






\bibitem{YMo16} Y. Mo, G. Tian, R. Car, V. N. Staroverov, G. E. Scuseria, and J. Tao, Phys. Rev. B 95, 035118 (2017).

\bibitem{YMoCPL} Y. Mo, G. Tian and J. Tao, Chem. Phys. Lett. {\bf 628}, 38$-$42 (2017).

\bibitem{mapkw} A. E. Mattsson, R. Armiento, J. Paier, G. Kresse, J. M. Wills, and T. R. Mattsson, J. Chem. Phys. {\bf 128}, 084714 (2008).

\bibitem{smrkp} J. Sun, M. Marsman, A. Ruzsinszky, G. Kresse, and J. P. Perdew Phys. Rev. B {\bf 83}, 121410(R).

\bibitem{mgga-vasp} J. Sun, M. Marsman, G. I. Csonka, A. Ruzsinszky,
P. Hao, Y. S. Kim, G. Kresse, and J. P. Perdew, Phys. Rev. B {\bf 84}, 035117 (2011).

\bibitem{perdewpnas} A. Patra, J. E. Bates, J. Sun, and J. P. Perdew, Proceedings of the National Academy of Sciences, 114, 44, E9188-E9196 (2017).





\bibitem{PJanthon13} P. Janthon, S. M. Kozlov, F. Vines, J. Limtrakul, and F. Illas, J. Chem. Theory Comput. 9, 1631-1640 (2013). 

\bibitem{PJanthon14} P. Janthon, S. (Andy) Luo, S. M. Kozlov, F. Vines, J. Limtrakul, D. G. Truhlar,and F. Illas, J. Chem. Theory Comput. 10, 3832-3839 (2014).


\bibitem{PBlaha09} P. Haas, F. Tran, and P. Blaha, Phys. Rev. B {\bf 79}, 085104 (2009).

\bibitem{FTran16} F. Tran, J. Stelzl, and P. Blaha, J. Chem. Phys. {\bf 144}, 204120 (2016).

\bibitem{GCsonks09} G. I. Csonka, J. P. Perdew, A. Ruzsinszky, P. H. T. Philipsen, S. Lebègue, J. Paier, O. A. Vydrov, and J. G. Ángyán, Phys. Rev. B 79, 155107 (2009).


\bibitem{GKresse13} L. Schimka, R. Gaudoin, J. Klimes, M. Marsman, and G. Kresse, Phys Rev B 87, 214102 (2013).

\bibitem{PHao12} P. Hao, Y. Fang, J. Sun, G. I. Csonka, P. H. T. Philipsen, and J. P. Perdew, Phys. Rev. B {\bf 85}, 014111 (2012).

\bibitem{GZhang08}G. Zhang, A. M. Reilly, A. Tkatchenko, and M. Scheffler, New J. Phys. 20 063020 (2018).

\bibitem{SJana18} S. Jana, A. Patra, and P. Samal, (to be appeared), J. Chem. Phys. (2018).


\bibitem{paw1} P. E. Bl\"ochl, Phys. Rev. B, 50:17953, (1994).

\bibitem{paw2} G. Kresse and D. Joubert, Phys. Rev. 59 , 1758 (1999).


\bibitem{vasp1} G. Kresse and J. Hafner, Phys. Rev. B 47 , 558 (1993); ibid. 49 , 14 251 (1994).

\bibitem{vasp2} G. Kresse and J. FurthmAller, Comput. Mat. Sci. 6 , 15 (1996). 

\bibitem{vasp3} G. Kresse and J. FurthmAller, Phys. Rev. B 54 , 11 169 (1996). 

\bibitem{uspp} G. Kresse and J. Hafner, J. Phys.: Condens. Matt. {\bf 6}, 8245 (1994).

\bibitem{JTao10} J. Tao, J. P. Perdew, and A. Ruzsinszky, Phys. Rev. B 81, 233102
(2010).

\bibitem{EOS1} F. D. Murnaghan, Proc. Natl. Acad. Sci. U.S.A. {\bf 30}, 244 (1944).


\bibitem{EOS2} A. B. Alchagirov, J. P. Perdew, J. C. Boettger, R. C. Albers, and
C. Fiolhais, Phys. Rev. B 63, 224115 (2001).

\bibitem{EOS3} A. B. Alchagirov, J. P. Perdew, J. C. Boettger, R. C. Albers, and
C. Fiolhais, Phys. Rev. B 67, 026103 (2003).

\bibitem{EOS4} F. Birch, Phys. Rev. 71, 809 (1947).

 \end{thebibliography}
\end{document}